\documentclass[aps,prbbib,twocolumn,epsf]{revtex4}

\usepackage{graphicx}

\begin{document}
\draft

\title{Organic photovoltaic bulk heterojunctions with spatially varying composition.}
\author{Paul M. Haney$^{1}$}

\affiliation{$^{1}$Center for Nanoscale Science and Technology,
National Institute of Standards and Technology, Gaithersburg,
Maryland 20899-6202, USA }

\begin{abstract}
Models of organic bulk heterojunction photovoltaics which include
the effect of spatially varying composition of donor/acceptor
materials are developed and analyzed.  Analytic expressions for the
current-voltage relation in simplified cases show that the effect of
varying blend composition on charge transport is minimal.  Numerical
results for various blend compositions, including the experimentally
relevant composition of a donor-rich region near the cathode (a
``skin layer" of donor material),  show that the primary effect of
this variation on device performance derives from its effect on
photocharge generation.  The general relation between the geometry
of the blend and its effect on performance is given explicitly.  The
analysis shows that the effect of a skin layer on device performance
is small.
\end{abstract}

\maketitle

\section{introduction}

Photovoltaic devices consisting of two types of organic materials
(referred to as donor (D) and acceptor (A)) have attracted
considerable scientific interest in recent years.  Their operation
consists of the generation of an exciton in the donor molecule,
which is then disassociated into free carriers at the D-A interface
(the electron is transferred to the acceptor molecule's lowest
unoccupied molecular orbital (LUMO), leaving a hole in the donor
molecule's highest occupied molecular orbital (HOMO)). Carriers
which avoid recombination are then collected by contacts. The
geometry first studied consisted of single D and A layers, with a
single planar interface \cite{tang}. The resulting efficiencies were
low (1~\%), owing in part to the short exciton diffusion length (10
nm) - only excitons within this short distance from the interface
lead to free carriers. It was subsequently discovered that blending
D and A together throughout the device thickness led to increased
efficiencies \cite{yu}, now above 5~\% \cite{eff1,eff2,eff3}. This
increase in efficiency is attributed to an increase in D-A
interfacial area; carrier transport is sufficiently robust to the
disorder present in the blend to accommodate reasonable quantum
efficiencies.  If the organic blend is completely homogeneous, the
contacts on the device must be different in order to break spatial
symmetry and permit a nonzero short-circuit current in a preferred
direction.  The key difference between the contacts is their work
function: a lower (higher) work function ensures that the contact
preferentially collects and injects electrons (holes). Hence it is
understood that the cathode collects electrons, and the anode
collects holes.

A major thrust of experimental efforts has been to attain control
over blend morphology in order to optimize both exciton
disassociation and charge transport. Recent examples include using
nanoimprint lithography to control the structure of the
donor-acceptor molecules' interfacial profile \cite{morph1}, or
using a graded donor-acceptor blend to optimize both carrier
collection and transport \cite{holmes}.  A key challenge of
engineering blend morphology is the measurement and characterization
of the structure of the organic blend. Techniques for accomplishing
this include atomic force microscopy \cite{hamadani}, ellipsometry
\cite{germack}, and X-ray photoelectron spectroscopy \cite{xu}.
%,and electron tomography \cite{andersson}.
These techniques have revealed that typical methods for fabricating
devices lead to a layer of enhanced donor molecule density at the
cathode, which has been attributed to surface energy differences
between the active layer and other components \cite{xu}. This would
seem to present an impediment to good device performance: the
cathode collects electrons, but in its vicinity is mostly holes!
Nevertheless, internal quantum efficiencies of 90~\% have been
observed in these materials \cite{schilinksy}, indicating that
charge collection is still a relatively efficient process
\cite{germack}.

In this work, I theoretically study the effect of a nonuniform blend
on organic photovoltaic (OPV) device performance.  I employ a
drift-diffusion equation to describe electron and hole transport, a
field and temperature dependent generation and recombination rate
that captures the exciton physics, and the Poisson equation for
electrostatics.  To this model I add the effect of a spatially
varying effective density of states (EDOS) (note that ``density of
states" refers to the number of states per volume per unit energy,
whereas ``effective density of states" refers to the number of
states per volume). Part I describes details of the model. In part
II, I present analytic solutions for the transport under certain
approximations; these point to the fact that the effect of a
spatially varying EDOS on charge transport is small. In part III, I
present numerical results which indicate that the primary effect of
a spatially varying EDOS is on the charge generation and ensuing
$J_{\rm sc}$.  It is shown that this can be understood in terms of
the overall geometry of the composition profile.  I conclude that
since the skin layer near the cathode is geometrically small on the
scale of the device thickness, its effect on performance is
similarly small.

\section{Model}

The model used to describe the system is similar to that found in
Ref. \onlinecite{koster}.  Its basic equations are presented here in
dimensionless form.  Table I shows the variable scalings used.  The
dimensionless drift-diffusion/Poisson equations including a
spatially varying EDOS are given as \cite{fonash}:
\begin{eqnarray}
J_n &=& n \left(-\frac{\partial}{\partial x} V - \frac{\partial}{\partial x} \left[{\rm ln} N\right]\right) + \frac{\partial}{\partial x} n~, \nonumber  \\
J_p &=& f_{\mu} \left[p \left(-\frac{\partial}{\partial x} V + \frac{\partial}{\partial x} \left[{\rm ln} P\right]\right) - \frac{\partial}{\partial x} p\right]~, \nonumber  \\
-\frac{\partial}{\partial x}J_n  &=& \frac{1}{f_{\mu}} \frac{\partial}{\partial x}J_p = G-R~, \label{eq:dd} \\
\frac{\partial^2}{\partial x^2}V &=& p-n~,\label{eq:poisson}
\end{eqnarray}
where $f_\mu = \mu_h/\mu_e$ is the ratio of hole to electron
mobility, $G$ is the carrier density generation rate, and $R$ is the
recombination.  $N(x)$ and $P(x)$ are the spatially-dependent
electron and hole effective density of states, respectively. $n$ and
$N$ are related via: $n = N e^{-\left(E_c-E_{F,n}\right)/kT}$, where
$E_{F,n}$ is the electron quasi-Fermi level, $E_c$ is the conduction
band edge, and all quantities are position-dependent (the densities
are assumed to be such that the system is in a nondegenerate
regime). $p$ and $P$ are related similarly. $N(x)$ and $P(x)$ are
fixed material parameters, while $n$ and $p$ are system variables
that depend on applied voltage and illumination. For a single band
semiconductor, the effective density of states $N$ is given by
$\frac{1}{\sqrt{2}}\left(\frac{m^*_n k_{\rm
B}T}{\pi\hbar^2}\right)^{3/2}$, where $m^*_n$ is the effective
electron mass.  In the present context of organic materials, $N$ is
more properly understood as the number of HOMO states per unit
volume, and is proportional to the donor molecule density.

\begin{table}[h!b!p!]
\caption{Normalization to dimensionless variables.  In the below
$N_0$ is the characteristic density (typically chosen to be on the
order of $10^{-25}~{\rm m}^{-3}$) $D_n$ is the electron diffusivity,
$\epsilon$ is the dielectric constant of the organic blend, $q$ is
the magnitude of the electron charge, $T$ is the temperature, and
$k_{\rm B}$ is Boltzmann's constant.}
\begin{tabular}{|c|c|}
  \hline
  Quantity & Normalization \\\hline\hline
  density & $N_0$ \\
  position & $$ $\sqrt{\epsilon k_B T / (q^2 N_0)}\equiv x_0$\\
  charge current & $q D_{n} N_0/x_0$ \\
  electric potential & $k_{\rm B} T/q$\\
  rate density & $x_0^2/N_0 D_n$ \\
  \hline
  \end{tabular}
  \end{table}

The boundary conditions are given as:
\begin{eqnarray}
n\left(0\right) &=& N\left(0\right)e^{-E_g+\phi_L}, \nonumber\\
p\left(0\right) &=& P\left(0\right)e^{-\phi_L},\nonumber\\
~n\left(L\right) &=& N\left(L\right)e^{-\phi_R},\nonumber\\
~p\left(L\right) &=& P\left(L\right)e^{-E_g+\phi_R},\label{eq:bcond}
\end{eqnarray}
where $L$ is the device thickness (this represents placing the anode
at $x=0$, and the cathode at $x=L$).  $\phi_{L(R)}$ is the absolute
value of the difference between HOMO (LUMO) and left (right) contact
Fermi level.  The boundary condition for the Poisson equation is:
\begin{eqnarray}
V(L)-V(0) = \left(E_g - \phi_L - \phi_R\right) - V_{\rm A},
\end{eqnarray}
where $V_{\rm A}$ is the applied voltage (with the sign convention
above, $V_{\rm A}>0$ corresponds to forward bias).

I consider only bimolecular recombination, with (dimensionless)
form:
\begin{eqnarray}
R = \left(np - n_i^2\right)~,
\end{eqnarray}
where $n_i^2=n_0p_0$, and $n_0~(p_0)$ is the equilibrium electron
(hole) density.  The carrier generation rate density is taken to be
spatially uniform.  As described in Ref. \cite{koster}, adding the
exciton density as a system variable modifies the source term in Eq.
(\ref{eq:dd}):
\begin{eqnarray}
\left(G-R\right) \rightarrow \tilde{P}\times  G_0 -
\left(1-\tilde{P}\right)\times R,
\end{eqnarray}
where $G_0$ is the exciton density generation rate, and $\tilde{P}$
is a field and temperature dependent factor which represents the
probability for an exciton to disassociate into free electron and
hole \cite{koster,braun}.  The field and temperature dependence is
described by Braun's extension of Onsager's theory of ion
disassociation in electrolytes \cite{onsager}.

Charge recombination and generation also generally depend on the
donor and acceptor effective density of states.  The total source
term of Eq. (\ref{eq:dd}) (denoted here by $U(x)$) is therefore of
the generic form:
\begin{eqnarray}
U(x) &=& \tilde{P} \times G_0 \times g \left[N(x),P(x)\right] -
\nonumber
\\&&~~\left(1-\tilde{P}\right) \times R \times r \left[N(x),P(x)\right].
\label{eq:geq}
\end{eqnarray}

The appropriate forms for $g$ and $r$ depend on several factors,
such as the dependence of the optical absorption and D-A interface
area on relative D-A composition.

%These important issues have been addressed in previous work (ref);
\label{sec:linear}
\begin{figure}[h!]
\begin{center}
\vskip 0.2 cm
\includegraphics[width=3.5in]{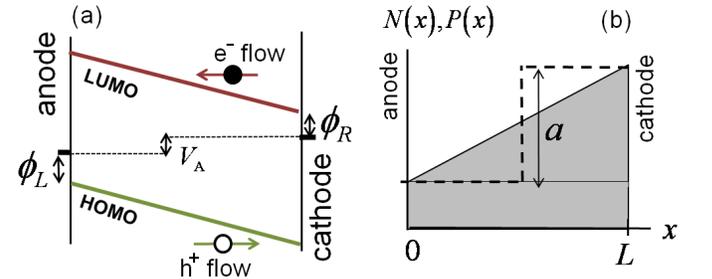}
\vskip 0.2 cm \caption{(a) Energy diagram for device model; cartoon
of particle flow depicts dark current in forward bias. (b) Spatial
dependence of EDOS: linear variation (shaded region), and step-like
change (dotted line) in both $N(x)$ and $P(x)$.}\label{fig:diagrams}
\end{center}
\end{figure}

\section{Analytic cases}
The set of equations described in Eq. (\ref{eq:dd}) can be solved
analytically for limiting cases, which can provide some insight into
the effect of a spatially varying EDOS on the transport.  Two cases
are considered here: the first is an exponentially varying EDOS
(which can be extrapolated to a linearly varying EDOS), and the
second is an abrupt, step-like change in the EDOS.  I present both
solutions first and discuss the physics they describe second.

In both cases the electric field $E$ is taken to be spatially
uniform (so that $V(x) = -E x$), and recombination is ignored. I
suppose further that $G$ is constant, and independent of $N,P$ (that
is, $g(N,P)=1$ in Eq. (\ref{eq:geq})).  The exponentially varying
EDOS is parameterized as:
\begin{eqnarray}
N\left(x\right) &=& P\left(x\right) =
A_0~e^{ax/L},\label{eq:expEDOS}
\end{eqnarray}
where $A_0 = a/\left(e^a-1\right)$ ensures that the total number of
states is independent of $a$.  Substituting the expressions for
electron (hole) current density $J_{n(p)}$ into the equation of
continuity (Eq. (\ref{eq:dd})) results in a second order
differential equation for the electron (hole) density $n$ ($p$). For
the EDOS of Eq. (\ref{eq:expEDOS}), the resulting general solution
is:
\begin{eqnarray}
n\left(x\right)&=&c_1 e^{\left(a-f\right)x}+c_2 +\frac{Gx}{a-f}~, \nonumber\\
p\left(x\right)&=&c_1 e^{\left(a+f\right)x}+c_2
+\frac{Gx}{a+f},\label{eq:npan}
\end{eqnarray}
where $c_1,~c_2$ are determined by the boundary conditions of Eq.
(\ref{eq:bcond}).  From this solution the current density can be
obtained directly.

I express the resulting current-voltage relation as a sum of dark
current and light current:
\begin{eqnarray}
J\left(V_{\rm A}\right) = J_D + G J_L~,
\end{eqnarray}
Both light and dark currents are well described by expanding to
lowest order in the spatial variation of EDOS parameter $a$; I take
$\phi_L=\phi_R=0$, and express the applied voltage dependence in
terms of $f = \left(E_g - qV_{\rm A}\right)/k_{\rm B}T$. $f$ is
bigger than 1 in the region of interest \cite{footnote1}, leading to
the further approximation that $\sinh f \approx \cosh f \approx 1/2
~e^f$.  It's useful to express current-voltage relation in terms of
that for a uniform EDOS and electric field:
\begin{eqnarray}
J_D^0 &=& \frac{2 f \left(e^{V_{\rm A}}-1\right)}{L\left(e^f-1\right)e^{V_{\rm A}}} \nonumber\\
J_L^0 &=& L\left(\frac{2}{f}-\coth \left(\frac{f}{2}\right)\right).
\end{eqnarray}

The dark and light current for the exponentially varying profile of
Eq. (\ref{eq:expEDOS}) is then found to be:
\begin{eqnarray}
J_D^{\rm exp} &\approx& J_D^0 \left( 1 + a^2\left(\frac{1}{12} -
\frac{1}{2f}\right) + O\left(a^4\right)+...\right)~, \nonumber\\
J_L^{\rm exp} &\approx& J_L^0 \left( 1 + a^2 \left(-\frac{2}{f^3} +
e^{-f}\right)+ O\left(a^4\right)+...\right).\label{eq:j1}
\end{eqnarray}

I next consider a step function form of $N(x),P(x)$. I use the
following form:
\begin{eqnarray}
N\left(x\right) = P\left(x\right) =\left\{
\begin{array}{rl}
1-a/2 & \text{if } x < L/2,\\
1+a/2 & \text{if } x \geq L/2.
\end{array} \right.
\end{eqnarray}
The general solutions for each region ($x<L/2,~x>L/2$) are of the
form given by Eq. (\ref{eq:npan}) with $a$=0.  In addition to the
boundary condition Eq. (\ref{eq:bcond}), there's an additional
boundary condition for this EDOS of continuity of charge and current
density at $x=L/2$.  Making the same approximations as above leads
to the following dark and light current:
\begin{eqnarray}
J_D^{\rm step} &\approx& J_D^0\left(1 - a^2 e^{-f/2}+O\left(a^4\right)+...\right)~,\nonumber \\
J_L^{\rm step} &\approx& J_L^0\left(1 - \frac{a^2}{2} \frac{f
e^{-f/2}}{f-2}+O\left(a^4\right)+...\right). \label{eq:j2}
\end{eqnarray}

A number of interesting and relevant features emerge from these
solutions: first, only even powers of $a$ appear in the expansions.
This is a consequence of the symmetry built in to the system: when
$\phi_L=\phi_R$ and $f_\mu=1$, electron particle transport from left
to right is equal to hole particle transport from right to left.  In
both the exponential and step-like cases above, holes encounter an
expansion in the EDOS along their transport direction, which
increases the hole current. Conversely, electrons encounter a
constriction, which decreases the electron current. To linear order
(and all odd orders) in the expasion/contraction parameter $a$,
these effects cancel each other so that the total charge current
only appears with even powers of $a$.  If the electron/hole symmetry
is broken, or the symmetry of the EDOS is reduced (by shifting the
step away from the center of the device), then odd powers of $a$ are
present (with prefactors whose magnitude reflects the degree of
symmetry breaking).

The other relevant feature of Eqs. (\ref{eq:j1}) and (\ref{eq:j2})
is the small magnitude of the $a^2$ prefactor.  Noting again that
$f$ is generally larger than 1 for the applied voltages of interest
to solar cells, it's clear by inspection that the prefactors are
much smaller than 1.  This indicates that the effect on transport of
a spatial variation of the EDOS is quite weak.

\begin{figure}[h!]
\vskip 0.2 cm
\includegraphics[width=2.5in]{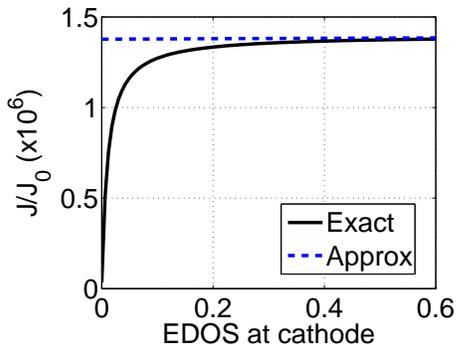}
\vskip 0.2 cm \caption{Extinction of current when EDOS goes to zero.
This is for the step-like change in EDOS, for parameters
$G=10^{-9},~V_{\rm A}=0.7$.  Both approximate and exact values are
shown (where the approximate expression is given by Eq.
(\ref{eq:j2})). It is seen that the current decreases substantially
only when the EDOS is nearly zero (or when $a$ is nearly 2).
}\label{fig:pinchoff}
\end{figure}

The intuitive picture that emerges from this analysis is that
electrons and holes can very easily ``squeeze" through regions of
reduced density.  A natural question concerns the way in which
transport is ultimately ``pinched off" by letting the density vanish
at a point in space.  This is shown in Fig. (\ref{fig:pinchoff}),
which shows the current in the step-like structure as $a\rightarrow
2$. The way in which the current vanishes is very steep; it is only
at very small values of EDOS at the cathode that the current drops
appreciably (in this limit, the approximation $a\ll 1$ used in
deriving Eq. (\ref{eq:j2}) is not satisfied, hence the discrepancy
between exact solution and Eq. (\ref{eq:j2})). However, for very
small values of HOMO and LUMO density in real systems, the model
presented here is likely not appropriate.  This point is discussed
more fully in the conclusion.

\section{numerical studies}

I next consider the effect of spatially varying EDOS when the
Poisson equation for the electric potential and bimolecular
recombination are included.  Recall that the dependence of the
generation and recombination on EDOS of electron $N$ and hole $P$ is
described generically as:
\begin{eqnarray}
\tilde{P}\times G_0\times g\left(N,P\right) -
\left(1-\tilde{P}\right) \times R  \times r\left(N,P\right)~.
\end{eqnarray}
I make the following ansatz for $g$ (the main conclusion can be
formulated in a way that's independent of this specific choice for
$g$):
\begin{eqnarray}
g\left(N,P\right) = P^2 N~.
\end{eqnarray}
This is motivated by the observation that the D-A interfacial area
requires both $P$ and $N$, hence $g$ has a factor of each; an extra
factor of $P$ is added since the exciton is initially generated in
the donor. $r$ is taken simply to be 1, since $R$ already has $N$
and $P$ dependence built in through $n$ and $p$.   Adding a factor
of $P$ to the recombination (so that the $N,P$-dependence of both
generation and recombination is the same) has only a weak effect on
the results.

\begin{figure}
\vskip 0.2 cm
\includegraphics[width=2.5in]{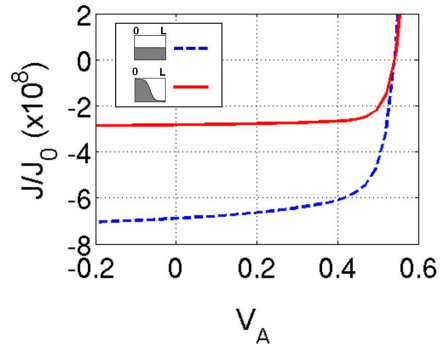}
\vskip 0.2 cm \caption{Current density-Voltage relation for two
spatial profiles of D-A EDOS profiles.  Blue dotted line is for
uniform EDOS profile, red line is for S-shaped EDOS profile, given
by Eq. (\ref{eq:tanh})}\label{fig:ivs}
\end{figure}

A range of composition profiles has been explored for the numeric
evaluation of device performance, and I present two representative
examples here:
\begin{eqnarray}
N_1(x)=1-P_1(x)&=&\frac{a}{\left(e^a-1\right)} e^{ax/L}~,\label{eq:nforms} \\
N_2(x)=1-P_2(x)&=&\frac{1}{2}\left( 1 + \left(1 - 2a\right)
\tanh\left(\frac{x-x_0}{\lambda}\right)\right)\label{eq:tanh}.\nonumber\\
\end{eqnarray}
%The way that the profile depends on $a$ for Eqs. (\ref{eq:nforms})
%and (\ref{eq:tanh}) is depicted above the plots in Fig.
%(\ref{fig:u}a) and (\ref{fig:u}b), respectively.
Fig. (\ref{fig:ivs}) shows the $J-V$ curves for the
$\left(N_2,P_2\right)$ case (Eq. (\ref{eq:tanh})) for the uniform
profile ($a=1/2$), and a sharp S-shaped profile ($a=0.95$).  Note
that the effect of the EDOS profile on the short circuit current
$J_{\rm sc}$ is substantial, while the effect on open circuit
voltage $V_{\rm oc}$ is small.

The previous analysis can explain the relative insensitivity of
$V_{\rm oc}$ to a nontrivial EDOS profile: the effect of a varying
EDOS profile on transport is weak, so that the injected current
required to offset the photogenerated current (and the corresponding
required voltage - $V_{\rm oc}$) is only weakly sensitive to changes
in EDOS \cite{footnote2}.

The change in $J_{\rm sc}$ can be understood as a direct consequence
of the model construction.  $J_{\rm sc}$ is the current collected in
the absence of an applied voltage, that is, in the absence of charge
injected from the contacts.  As such it is simply equal to the total
charge generation rate in the device: $J_{\rm sc} = \int dx
\left({\rm Generation}(x) -{\rm Recombination}(x)\right)$.  As
described above, this is directly parameterized as:
\begin{eqnarray}
J_{\rm sc} &=& \int dx \left( \tilde{P} \times G_0 \times g
\left[N(x),P(x)\right] - \right. \nonumber \\&&
\left.~~~~~~~~(1-\tilde{P}) \times R \times
r\left[N(x),P(x)\right]\right). \label{eq:jscintegral}
\end{eqnarray}
In analyzing the effect of $N(x), P(x)$ on $J_{\rm sc}$, it is
instructive to separate the $N,P$ dependence of the generation from
the above integral.  This leaves a quantity $\delta U$ which depends
only on the geometry of the D-A EDOS profile:
\begin{eqnarray}
\delta U &=& \int dx ~g\left[N(x),P(x)\right] \label{eq:dU}\\
&=& \int dx ~N(x)   P^2(x). \label{eq:dUmine}
\end{eqnarray}

Strictly speaking the integral in Eq. (\ref{eq:jscintegral}) does
not factorize in  a manner that leads lead directly to a $\delta U$
term.  However, as I show in the following, $\delta U$ is a good
predictor of the effect of the geometry of the EDOS on the device
performance.

\begin{figure}[h!]
\vskip 0.2 cm
\includegraphics[width=3in]{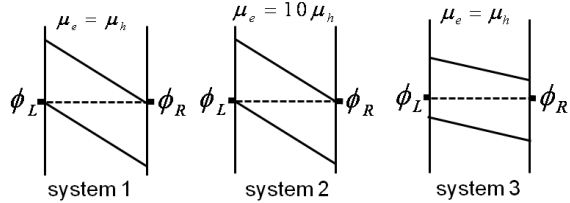}
\vskip 0.2 cm \caption{Cartoon of the three system
parameterizations: system 1: $\phi_L=\phi_R=0,~\mu_h=\mu_e$, system
2: $\phi_L=\phi_R=0,~\mu_h=10\mu_e$, system 3:
$\phi_L=\phi_R=0.2~{\rm eV},~\mu_h=\mu_e$}\label{fig:systems}
\end{figure}

For each EDOS profile, I also vary other system parameters.  The
three different parameterizations are shown in Fig.
(\ref{fig:systems}).  In system 1, HOMO/LUMO levels are aligned with
cathode/anode Fermi levels ($\phi_L=\phi_R=0$), and electron and
hole mobility are equal.  For system 2, $\phi_L=\phi_R=0$, but
electron and hole mobilities are not equal ($\mu_e = 10\mu_h$). In
system 3, the HOMO/LUMO are offset from cathode/anode by $0.2~{\rm
eV}$ ($\phi_L=\phi_R=0.2~{\rm eV}$), and electron/hole mobilities
are equal.

\begin{widetext}

\begin{figure}[h!]
\vskip 0.2 cm
\includegraphics[width=6.75in]{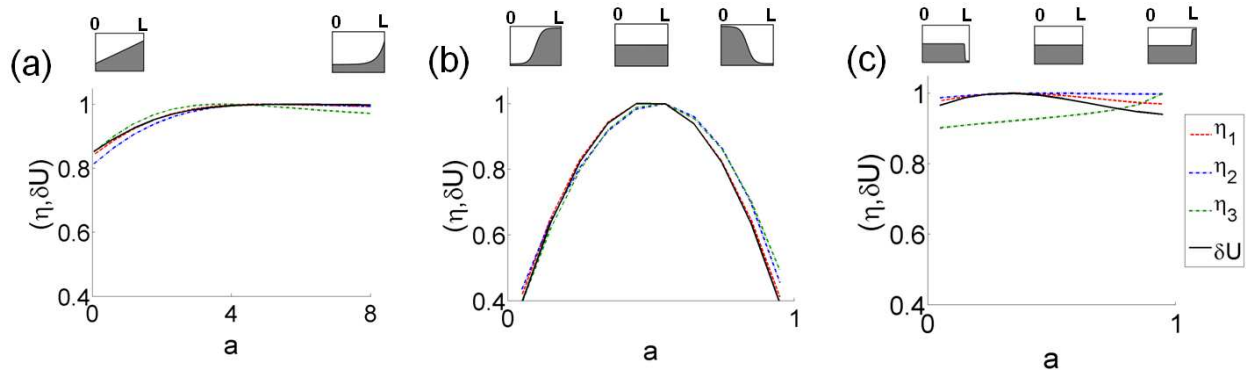}
\vskip 0.2 cm \caption{The efficiency $\eta$ and geometrical factor
$\delta U$ (normalized by their maximum value) versus geometrical
parameter $a$ for (a) exponentially varying profile ($N_1(x)$ of Eq.
(\ref{eq:nforms})) (b) S-shaped profile $N_2(x)$ of Eq.
(\ref{eq:tanh}), with $x_0=L/2$, $\lambda=L/8$), (c) ``skin" layer
geometry.  Representations of the spatial variation of EDOS as a
function of $a$ are shown above the figure.  The gray and white
regions represent $N(x)$ and $P(x)$, respectively.  The efficiency
closely follows the geometrical factor $\delta U$ for most cases.
For each geometry I use the three system parameterizations described
in Fig. (\ref{fig:systems}) (the subscript of $\eta$ specifies the
system parameterization). }\label{fig:u}
\end{figure}

\end{widetext}

Fig. (\ref{fig:u}a) and (\ref{fig:u}b) shows $\delta U$ as the
profile parameter $a$ is varied, for various EDOS configurations
given by Eqs. (\ref{eq:nforms}) and (\ref{eq:tanh}), respectively.
This is shown for the three system parameterizations.  The overall
device efficiency $\eta$ tracks $\delta U$ very closely for all of
these cases (the efficiency is proportional to the maximum absolute
value of $(J V)$ in the 4th quadrant of the $J-V$ plane). For this
reason I conclude that the primary effect of a spatially varying
EDOS on device performance is to change the total carrier generation
rate and ensuing $J_{\rm sc}$. $\delta U$ in Fig. (\ref{fig:u}) is
calculated using Eq. (\ref{eq:dUmine}), however the conclusion is
valid for any choice of $g$ I've tried. Hence the effect of a
nonuniform blend on performance can be approximately specified in
the generic form given by Eq. (\ref{eq:dU}).

Next I turn to the experimentally motivated geometry of a skin layer
of D near the cathode.  It's parameterized as:
\begin{eqnarray}
N(x)=1-P(x)=\frac{2+a}{4} + \frac{2-a}{4}
\tanh\left(\frac{x-x_0}{\lambda}\right),\nonumber\\
\end{eqnarray}
with $\lambda = 0.0075~L$, $x_0=0.05~L$.  Fig. (\ref{fig:u}c) shows
how the efficiency evolves as the skin layer goes from mostly D-like
(small $a$), to an even D-A mix, to mostly A-like (large $a$) (the
experimentally realistic case is smaller $a$). The change in
efficiency is a rather small effect for all three cases (a maximum
of 10~\% change). Also shown is the geometrical factor $\delta U$
(solid line).  The efficiency of system 1 conforms most closely to
the geometrical factor profile dependence. Inspection of the $J-V$
curves for the three systems reveals subtle differences in the
fill-factor between the three; there is no simple or obvious source
for the difference in behavior between the three system
parameterizations. The difference in behavior between the three
systems is more conspicuous for the skin layer geometry because the
effect of a nonuniform blend is smaller for the skin layer, so that
the overall performance is more sensitive to other system
parameters.  (When the blend profile leads to larger effects, for
example that shown in Fig. (\ref{fig:u}b), there is a similar
dependence of device performance on profile for all system
parameterizations.) Nevertheless, the important conclusion common to
all three system parameterizations of the skin layer geometry is
that the effect of the skin layer is small. Its smallness can be
understood in terms of the analysis of the previous sections.  The
analytic work points to the fact that the effect of blend
non-uniformity on charge transport is generically small (except in
extreme cases). The numerical work of the previous test cases
indicates that the effect of blend non-uniformity can be understood
in terms of its effect on charge generation and resulting $J_{\rm
sc}$ - and that this effect is essentially geometrical (see Eq.
(\ref{eq:dU})). Since a skin layer is by definition geometrically
small, its effect is similarly small.

\section{Conclusion}

In this work I presented a simple model for the effect of nonuniform
blend profiles on OPV device performance.  The main effect of a
nonuniform D-A blend is on the the charge generation and resulting
short-circuit current: in regions where the blend is primarily of
one type at the expense of the other, there is less charge
generation due to a reduced D-A interfacial area. The details of how
charge generation depends on local blend mix are complicated, and
involve almost all aspects of OPV device operation (e.g optics
\cite{moule}, exciton diffusion \cite{holmes}, etc.). The influence
of a nonuniform blend on electron and hole transport is a weaker
effect.

It's important to appreciate the simplicity of the model presented
here relative to the complexity of real OPV devices.  Two
simplifications of the model are: its treatment of the metal-organic
interface, and its restriction to 1 spatial dimension. I make no
attempt to capture the effect of a skin layer geometry on the
metal-organic contact.  The physics at this interface is included
most simply as a finite recombination velocity \cite{scott} (which
can also depend on temperature and field \cite{lacic}).  A hallmark
of less effective charge collection/injection at this interface is
S-shaped $J-V$ curves \cite{deibel}.  This feature is correlated to
metal contact deposition techniques \cite{deibel}, and is not
ubiquitously observed in devices. I therefore conclude that the
details of the metal-organic contact is not directly tied to the
phase segregation in the organic blend.

A more severe approximation of this model is its restriction to 1-d.
When the EDOS is small, the charge and current density is also
small.  However, experiments reveal localized hot-spots of
conducting paths \cite{hamadani}.  A 1-d model necessarily averages
these localized hot-spots or large current density over the entire
cross-sectional area, leading to a diffuse current.  As the overall
area of hot-spots decreases, the charge and current density they
must accommodate increases, and current may become space-charge
limited.  A 1-d model is unable to capture the physics described in
this scenario.  However for less extreme cases, the treatment
described here offers the simplest account for a spatially varying
blend structure.

I acknowledge very useful discussions with Behrang Hamadani and Lee
Richter.


\begin{thebibliography}{00}
\bibitem{footnote1}
The most relevant region of the current-voltage relation is in the
4th quadrant.  In this region, $V_{\rm A}< E_g$; the different
between $E_g$ and $qV_{\rm A}$ is scaled by $1/k_{\rm B} T = 40
\left({\rm eV}\right)^{-1}$, so that $f$ is generally much larger
than 1.
\bibitem{footnote2}
The change in $J_{\rm sc}$ induced by a spatially varying EDOS has
some effect on $V_{\rm oc}$ as well, but this is also a small
effect, as $V_{\rm oc}$ generically varies only logarithmically with
$J_{\rm sc}$.
\end{thebibliography}
\end{document}